# Towards the Implementation of Smart Homes

**Nesreen Mufid, EiZ Engineering, Amman, Jordan. Email: nesreen@eiz-eng.com.jo**

## Introduction

The use of automation technology has become an essential aspect of our daily routines, encompassing our personal space, means of transportation, and work environments. With the availability of diverse communication interfaces and standards, the process of establishing monitoring and automation systems has become a simple and productive endeavor [1]. One of the most noteworthy examples of this is the smart home, which is essentially a home that leverages automation through different communication technologies, enabling homeowners to monitor and control their home appliances and security locks remotely [2]. To achieve this, the smart home is equipped with a complete network of electrical devices, including sensors and actuators that communicate seamlessly, allowing homeowners to manage their homes from afar [3]. With this capability, homeowners can control various aspects of their homes, such as lighting, air conditioning, and heating, from wherever they are. This technology provides the added benefits of comfort, luxury, energy savings, and security, making the home environment even more appealing [4]. The beauty of home automation is that it's flexible, allowing people to monitor and control their homes remotely from anywhere. This system also notifies homeowners of the status of their homes while they are away, through smartphone or tablet connectivity [5]. In recent years, the popularity of smart homes has surged due to the rise of the internet, which has made using smartphones and computers more accessible and affordable to people.

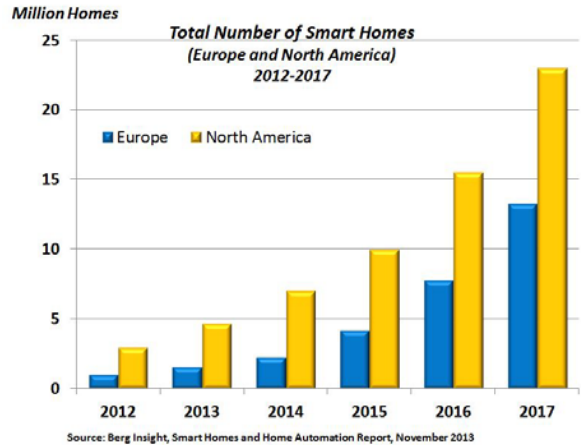

*Figure 1-1 number of present and future of smart homes in North America and Europe from 2012 to 2017 [3]*

However, the studies show that technology of home automation itself has not been globally adopted, even though it was available to consumers during the past three decades. This adaptation failure referred to different reasons such as, inflexibility, lack of manageability, difficulties to achieve security at home [5]. In addition to that, the cost of different electrical devices such as sensors, video cameras and programmable lighting was expensive in the past years. Also, some of the other sensors (e.g. motion and vibrating) were in the process of gaining commercial acceptance. Nevertheless, in the past few years, the smart home market demand is increasing due to the affordably for average consumers. A recent study by ABI research in (October2013) shows and predicts the growth of total number of smart homes in Both Europe and North America as shown in figure 1.1 and figure 1.2 [6].

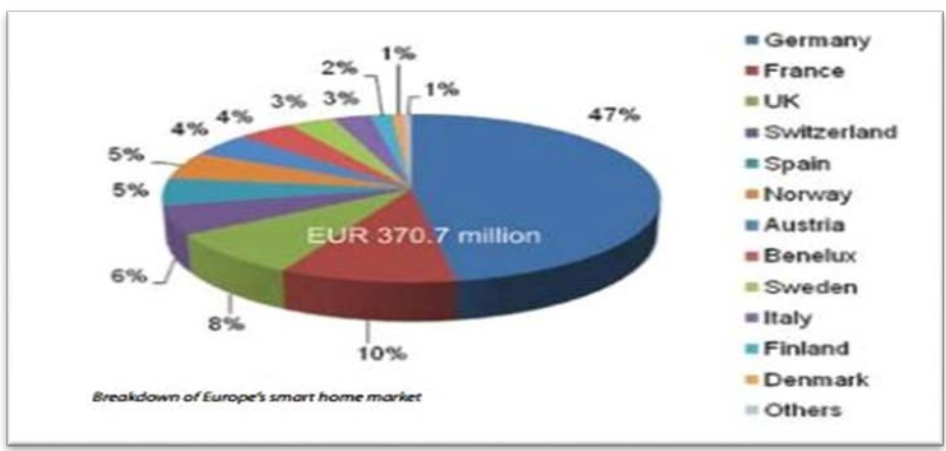

*Figure 1-2 Percentage of distribution of smart home products in European countries (this is the distribution in Europe among its countries and there is no prediction as u say above). [6]*

## 1.1. Background

In the era of 1960s to 1970s, scientists and researchers had visions about robots, artificial intelligence and electronic brains and the future seemed to be promising. In addition to that, women made careers outside their home and started to spend less time inside it [7]. New technologies in home appliances such as, central heating and food processors were appearing in the market. Also, partial degree of automation emerged on home environment. In year 1969, a project called ARPANET took place in USA, became operational [8]. It was the first packed switching network in the world, which later evolved to be the internet. This project was essential at that time, because computers were expensive and researchers looked for approach to exchange data and information. After the invention of microcontrollers, home automation the technologies were adopted by the construction services and appliance manufacturers [9]. Later on 1990s period, when computers gave a huge role in communications, different smart homes projects have started. Then at the 1998, came the intelligent room project, which was implemented by MIT. This room had the automation concept, containing a computer vision and speech and motion identification system. In addition to that, the cameras were used to recognize and trace people, while the speech recognition used for monitoring and getting auditory response. A picture of that intelligent room in MIT is shown in Figure 1.3 [10].

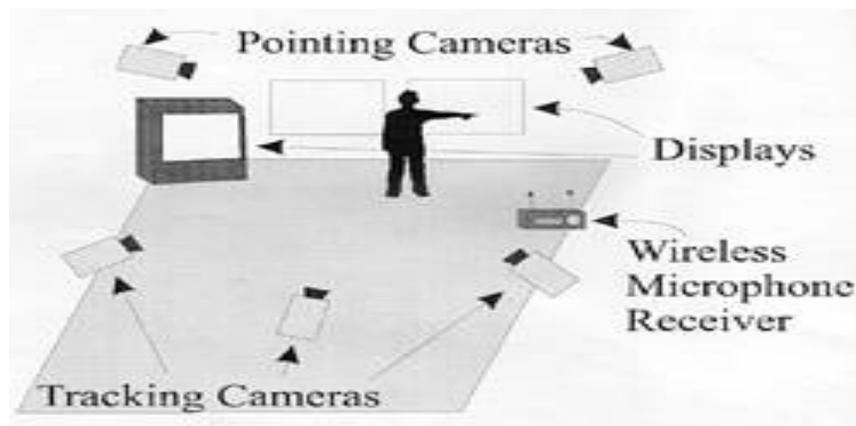

Figure 1-3 Brief description of the first intelligent room was in MIT in 1998 [10]

Finally starting in 2000s, time saving home appliances came to the market, which known as a smart device. Those devices are modern technological sensors that developed to be compatible with different applications. Such examples for those sensors are, gas detector sensors, motion sensors and smoke sensors. Modern smart homes were built in the basis of controlling those multi sensors that provide different tasks inside the environment of a place called home. A picture of the evolution of smart home through history is shown in figure 1.4 [11].

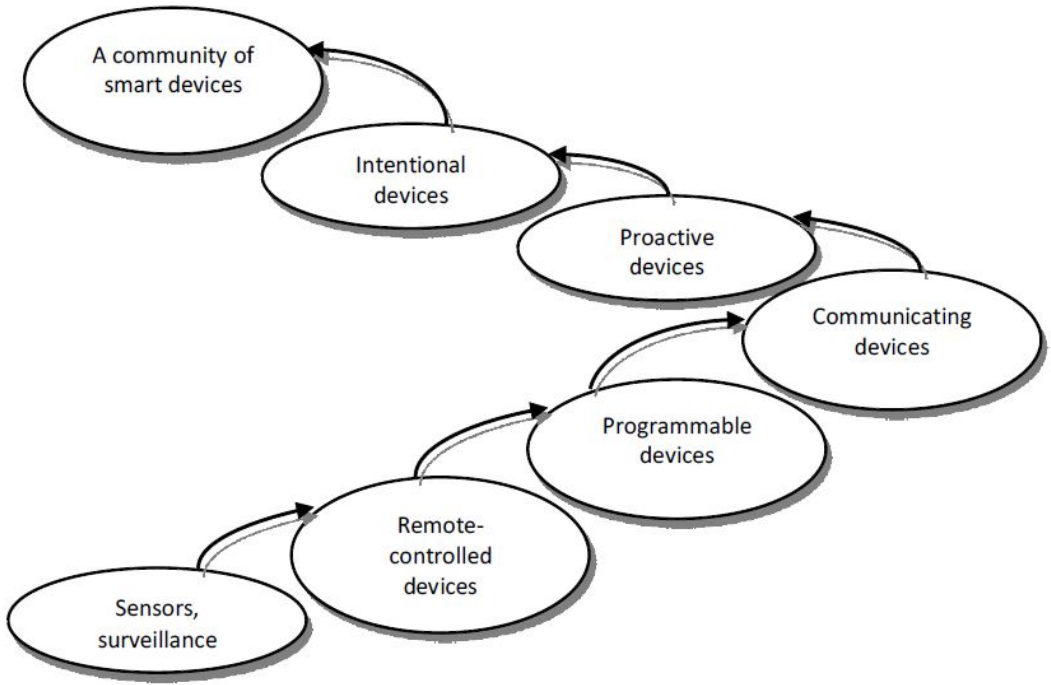

Figure 1-4 Evolution of the smart home concept [11]

## 1.2. Previous work

Smart home projects that exist around the world, share the same principle and concepts. Commonly, a smart home contains network of sensors and actuators that connect to the home appliances and send a signal to a controlling unit wirelessly to inform the user. In 2001, a telecommunication Norwegian company called Telenor designed a complete smart home. It was designed for demonstration purposes to give the people an idea and a complete vision of a modern smart home. The design focused on the communication technology to monitor the home and support the family members when they are moving from one room to another, as shown in Figure 1.5.

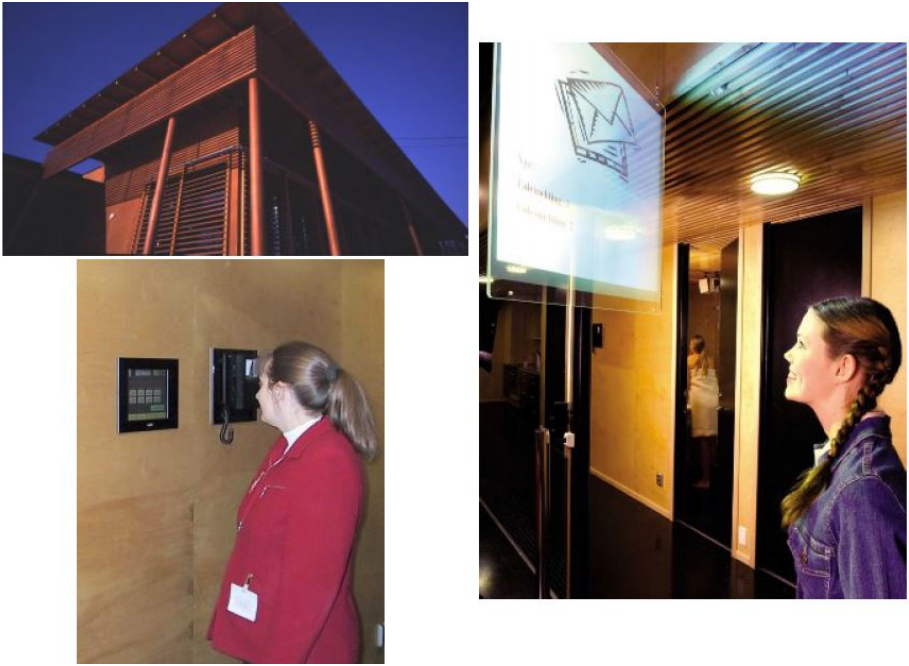

Figure 1-5 A complete smart home design by the Norwegian Company Telenor in 2001.

    In 2002 Philips research designed an intelligent ambient lab called (Philips Home laboratory). This lab was a fully equipped home, devoted to test different prototypes with the available technology [12]. In addition to that, the lab was built with a variety of features such as, user position, communication recognition and different screens for displaying information. The idea was to conduct several tests among family members and teenagers, to discover new approaches of interaction by sharing digital media. The automation approaches were used in the lab in order to monitor the equipment are, using gestures and speech commands. Different automation technologies tested by Philips research are now available in the market. A picture of Philips home is shown in Figure 1.6 [13].

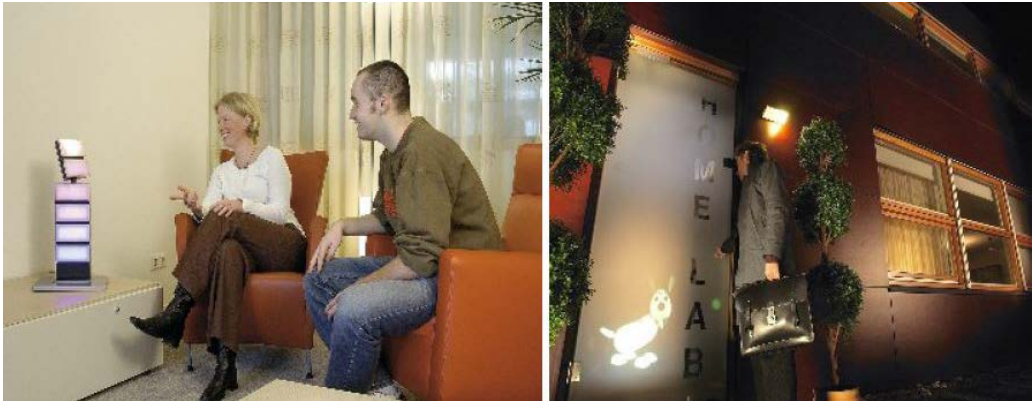

Figure 1-6 Philips home lab, built in Netherlands in 2004 [13]

In 2000, Georgia-Tech in Atlanta completed a project called "The Aware Home". Mainly, testing and evolution analysis of technology performed on it, which linked to the future living. The concept of the Aware home implies from its name, the home should be aware of the home members when they are inside it and aware of their activities. Based on that concept, tracking sensors are placed inside the home in order to monitor the movements of users and identify them easily. A picture of the aware home is shown in Figure 1.7 [14].

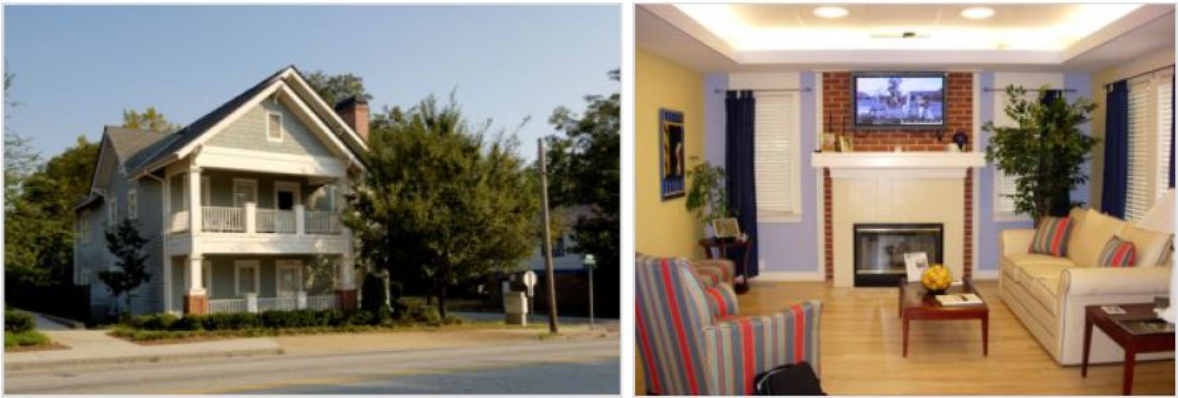

Figure 1-7 Aware Home, built in Atlanta in 2000 [14]

Home automation [15] refers to the application of M2M communication methodology to enable remote and automatic control of basic domestic activities within the home [16]. This project is centered around developing a cost-effective smart home equipped with various sensors that perform different functions within the home. These functions include remotely monitoring and controlling the lighting system by switching lights on and off, remotely automating home appliances, and remotely controlling the security system [17]. These features allow users to stay informed about the status of their homes even when they are away. The system relies on enabling M2M communication [18] via the cellular network infrastructure [19], which offers mobility and flexibility for users, allowing them to move around the country or even the world through roaming services. Additionally, this network enables a large number of M2M devices, such as sensors [20], to be efficiently connected. Unlike short-range M2M communication standards like Bluetooth [21] or ZigBee [22], cellular networks offer long-distance coverage, allowing users to move freely. The four main objectives of remotely automating a smart home are:

1. Security monitoring [23].

2. Home appliances remote control [24].

3. Energy saving through the optimization of the appliances turn ON/OFF times [25].

4. Safety monitoring [26].

**References**

[1] Michael Margolis, Arduino Cookbook, 2nd ed., Shawn Wallace and Brian Jepson, Eds. Newyork, USA: O'Reilly Media, Inc, 2012.

[2] Vini Madan and S.R.N, "GSM-Bluetooth based Remote Monitoring and Control System with Automatic Light Controller ," International Journal of Computer Applications , vol. 46, no. 0975 - 8887, May 2012.

[3] Dale Cigoy, "How to Selcet the Right Temperature," Keithyley Instruments, Inc., Cleveland, Electrical 2816, 2007.

[4] Toril Laberg, Haakon Aspelund, and Hilde Thygesen, "SMART HOME TECHNOLOGY, Planning and management in municipal services," Directorate for Social and Health Affairs, the Delta Centre, Oslo, 2005.

[5] N. Zorba and A.I. Pérez-Neira, "Opportunistic Grassmannian Beamforming for Multiuser and Multiantenna Downlink Communications", IEEE Transactions on Wireless Communications, no. 4, April 2008.

[6] Ali Isilak, "SMART HOME APPLICATIONS FOR DISABLED PEOPLE BY USING WIRELESS SENSOR NETWORK ," Department of Computer Engineering, Yeditepe University, Faculty of Engineering and Architecture , Engineering Project 2010.

[7] N. Zorba and A.I. Pérez-Neira, "Robust Power Allocation Schemes for Multibeam Opportunistic Transmission Strategies Under Quality of Service Constraints", IEEE JSAC special issue on MIMO for Next-Generation Wireless Networks, no.8, August 2008.

[8] E. Kartsakli et.al., "A Threshold-Selective Multiuser Downlink MAC scheme for 802.11n Wireless Networks", IEEE Transactions on Wireless Communications, April 2011.

[9] R. Imran et.al., "Quality of Experience for Spatial Cognitive Systems within Multiple Antenna Scenarios", IEEE Transactions on Wireless Communications, vol. 12, no.8, August 2013.

[10] Adiline Macriga, "Smart Home Monitoring and Controlling System Using Android Phone," International Journal of Emerging Technology and Advanced Engineering, vol. 3, no. 11, pp. 426,427, November 2013.

[11] E. Datsika et.al., "Cross-Network Performance Analysis of Network Coding Aided Cooperative Outband D2D Communications", IEEE Transactions on Wireless Communications vol. 16, May 2017.

[12] Alan G. Smith, Introduction to Arduino., 2011.

[13] B. Khalfi et.al., "Efficient Spectrum Availability Information Recovery for Wideband DSA Networks: A Weighted Compressive Sampling Approach", IEEE Transactions on Wireless Communications, vol. 17, April 2018.

[14] Meensika Sripan, Xuanxia Lin, Ponchan Petchlorlean, and Mahasak Ketcham, "Research and Thinking of Smart Home Technology," in International Conference on Systems and Electronic Engineering, Thailand, 2012.

[15] Z. Wu et.al., "Device-to-Device Communications at the TeraHertz band: Open Challenges for Realistic Implementation", IEEE Communications Standards Magazine, 7 (1), 82-87, December 2022.

[16] A.Z. Abyaneh et.al., "Empowering Next-Generation IoT WLANs Through Blockchain and 802.11 ax Technologies", IEEE Transactions on Intelligent Transportation Systems, August 2022.

[17] M. Alzard et.al., "Performance Analysis of Resource Allocation in THz-based Subcarrier Index Modulation Systems for Mobile Users", IEEE Access 9, 129771-129781, September 2021.

[18] A. Elbery et.al., "IoT-Based Crowd Management Framework for Departure Control and Navigation", IEEE Transactions on Vehicular Technology 70 (1), 95-106, January 2021.

[19] B. Hamdaoui et.al., "IoTShare: A Blockchain-Enabled IoT Resource Sharing On-Demand Protocol for Smart City Situation-Awareness Applications", IEEE IoT Journal 7 (10), 10548-10561, November 2020.

[20] B. Hamdaoui et.al., "Dynamic Spectrum Sharing in the Age of Millimeter Wave Spectrum Access", IEEE Network 34 (5), 164-170, July 2020.

[21] A. El-Wakeel et.al., "Robust Positioning for Road Information Services in Challenging Environments", IEEE Sensors Journal 20 (6), 3182-3195, December 2019.

[22] A. El-Wakeel et.al., "Towards a Practical Crowdsensing System for Road Surface Conditions Monitoring", IEEE Internet of Things Journal 5 (6), 4672-4685, March 2018.

[23] E. Datsika et.al., "Software Defined Network Service Chaining for OTT Service Providers in 5G Networks", IEEE Communications Magazine, 55 (11), 124-131, November 2017.

[24] S. Alanazi et.al., "Reducing Data Center Energy Consumption Through Peak Shaving and Locked-in Energy Avoidance", IEEE Transactions on Green Communications and Networking 1 (4), 551-562, August 2017.

[25] B. Khalfi et.al., "Optimizing Joint Data and Power Transfer in Energy Harvesting Multiuser Wireless Networks", IEEE Transactions on Vehicular Technology 66 (12), 10989-11000, June 2017.